# Tests of a Digital Hadron Calorimeter

Burak Bilki[1], John Butler[2], Ed May[3], Georgios Mavromanolakis[4,•], Edwin Norbeck[1], José Repond[3,*], David Underwood[3], Lei Xia[3], Qingmin Zhang[3,#]

1 – University of Iowa, Iowa City, IA 52242-1479, U.S.A.

2 – Boston University, 590 Commonwealth Avenue, Boston, MA 02215, U.S.A.

3 – Argonne National Laboratory, 9700 S. Cass Avenue, Argonne, IL 60439, U.S.A.

4 – Fermilab, P.O. Box 500, Batavia, IL 60510-0500, U.S.A.

In the context of developing a hadron calorimeter with extremely fine granularity for the application of Particle Flow Algorithms to the measurement of jet energies at a future lepton collider, we report on extensive tests of a small scale prototype calorimeter. The calorimeter contained up to 10 layers of Resistive Plate Chambers (RPCs) with 2560 1×1 $cm^2$ readout pads, interleaved with steel absorber plates. The tests included both long-term Cosmic Ray data taking and measurements in particle beams, where the response to broadband muons and to pions and positrons with energies in the range of 1 – 16 GeV was established. Detailed measurements of the chambers efficiency as function of beam intensity have also been performed using 120 GeV protons at varying intensity. The data are compared to simulations based on GEANT4 and to analytical calculations of the rate limitations.

## 1  Digital Hadron Calorimetry

To fully exploit the physics potential of a future lepton collider, operating at center-of-mass energies of 0.5 TeV and above, requires an unprecedented jet energy resolution, of the order of a factor of two better than previously attained. The preferred approach to achieve this performance is through the application of Particle Flow Algorithms (PFAs) [1]. FPAs utilize the tracker to measure the momenta of charged particle, the electromagnetic calorimeter to measure the energy of photons and the combined electromagnetic and hadronic calorimeters to measure the energy of neutral hadrons, i.e. the neutrons and $K_L^0$'s. The major challenge of PFAs is the association of energy deposits in the calorimeter to the charged or neutral particles originating from the interaction point. This challenge is met with the development of highly segmented or so-called imaging calorimeters [2].

In this context we present tests of a Digital Hadron Calorimeter (DHCAL) utilizing Resistive Plate Chambers (RPCs) as active medium. Our RPCs are based on two different designs [3] with either two or only a single glass plate. In the latter case the readout board serves as anode and closes the gas volume. The chambers use 1.1 mm thick glass as resistive plates and are operated in avalanche mode. The gas gap is 1.2 mm and is maintained with the help of fishing lines spaced 5 cm apart.

---

• Also affiliated with University of Cambridge, Cavendish Laboratory, Cambridge CB3 OHE, U.K.
* Corresponding author: repond@hep.anl.gov
# Also affiliated with Institute of High Energy Physics, Chinese Academy of Sciences, Beijing 100049, China and Graduate University of the Chinese Academy of Sciences, Beijing 100049, China.

The chambers are read out with a readout board containing 1×1 cm$^2$ pads and a binary (or digital) electronic readout system [4]. The latter is based on the DCAL chip [5], which was developed by Argonne and Fermilab and which has been optimized for the readout of large number of channels.

The project is being carried out by the DHCAL collaboration [6] including 36 people from Argonne National Laboratory, Boston University, Fermilab, Iowa University and the University of Texas at Arlington. It is part of the scientific program of the international CALICE collaboration [7], which develops finely segmented calorimetry for future lepton colliders.

## 2   The Vertical Slice Test

In preparation for the construction of a large size prototype calorimeter [8], the DHCAL collaboration assembled a small scale prototype calorimeter. This calorimeter utilized the entire readout chain, as envisaged for the large scale prototype, and so was called a Vertical Slice Test (VST). The VST included up to 10 RPCs, each with an area of 20×20 cm$^2$, and 16×16 cm$^2$ readout boards, each with 256 readout channels. The RPCs were based on the default design with two glass plates, except for one chamber which featured the single glass plate design.

The prototype was extensively tested with Cosmic Rays and in the Fermilab test beam [9]. In the following we shall briefly summarize the major results.

## 3   RPC Performance Tests

The performance tests include measurements of the noise rate, the detection efficiency for minimum ionizing particles (MIPs) and the pad multiplicity, i.e. the average number of pads firing in events where a MIP has fired at least one pad.

### 3.1   Measurement of the environmental dependence of the RPC performance

The performance criteria of a stack of RPCs were correlated with measurements of the ambient temperature, the atmospheric pressure and the humidity of the surrounding air [10,11]. Table 1 summarizes the results for 2-glass RPCs. The chambers were operated within the efficiency plateau, hence only small changes in efficiency were observed. As expected no evidence for any dependence on the humidity of the surrounding air was discovered.

| Performance variables | Changes for $\Delta T = 1^0$C [%] | Changes for $\Delta p = 100$ Pa [%] |
|---|---|---|
| Noise rate | 14 ± 2 | 0.02 ± 0.69 |
| Efficiency | 0.26 ± 0.05 | 0.06 ± 0.01 |
| Pad multiplicity | 2.0 ± 0.1 | 0.30 ± 0.02 |

Table 1: Summary of environmental dependences of the performance of 2-glass RPCs

Of particular interest is the study of the performance as function of gas flow. Figure 1 shows a steep increase in noise rate and pad multiplicity for gas flow rates below 0.3 cm$^3$/minute, which corresponds to about eight volume changes per day. The reasons for this degradation in performance are not yet fully understood, but are subject of further investigations.

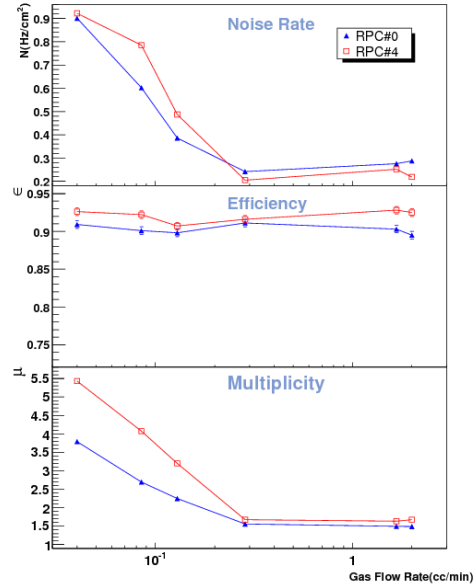

Figure 1: Performance as function of gas flow for 2-glass RPCs

.

### 3.2 Measurement of the rate capability of RPCs

The MIP detection efficiency as function of beam intensity was studied [12] with 120 GeV protons, as provided by the Fermilab test beam. For rates in excess of 100 Hz/cm$^2$ a clear loss in efficiency was observed. As function of time after the beginning of a beam spill the loss can be parameterized by the sum of an exponential and a constant. To understand this effect, we developed an analytical calculation based on the assumption that the loss of efficiency is due to the effective voltage drop caused by the current through the chamber. The calculation is found to reproduce the observed functional form of the efficiency drop as a function of spill time. Figure 2 compares the measured efficiency at the end of the spill with the results of the calculation. Good agreement is observed.

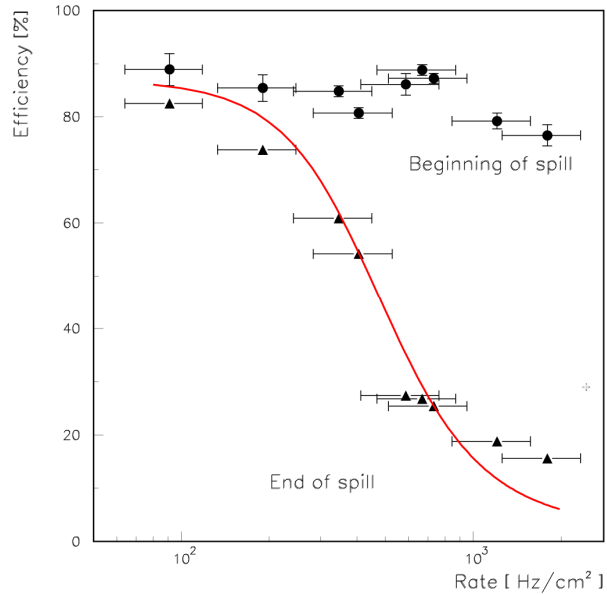

Figure 2: MIP detection efficiency at the beginning and at the end of a spill of 120 GeV protons. The red line is the result of an analytical calculation of the efficiency at the end of the spill.

## 4 Tests of the calorimeter prototype

The tests included six chambers interleaved with absorber plates containing a combination of a 16 mm thick stainless steel and a 4 mm thick copper plate. The stack was exposed to a broadband muon beam and to positrons and pions in the energy range of 1 – 16 GeV from the Fermilab test beam.

### 4.1 Simulation strategy

The set-up was simulated with a GEANT4 based program linked to a detailed simulation of the response of RPCs by a standalone program called RPC_sim [13]. The RPC_sim program generates a signal charge Q, distributes this charge over the pads, sums up all charges on a given pad, and applies a threshold T to identify the pads with hits. The response is controlled by four parameters of which three were tuned to describe the response to muons and the forth was tuned to describe the 8 GeV positron data. The pion data is subsequently simulated without further adjustment of these performance parameters.

### 4.2 Measurement with Muons

Broadband muons were obtained from the 120 GeV primary proton beam and a 3 m beam blocker. The muons were used to measure the average efficiency and pad multiplicity of the chambers under varying operating conditions (high voltage and threshold setting) [14]. Figure 3 shows the measured average pad multiplicity as a function of efficiency. The latter was varied applying different high voltage and threshold settings. Note, that the 1-glass design exhibits a constant value of the pad multiplicity close to unity and independent of the efficiency.

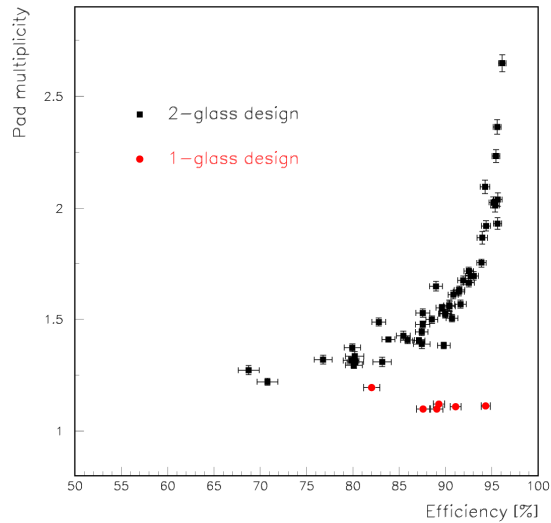

Figure 3: Average pad multiplicity versus efficiency for 2-glass (black dots) chambers and the 1-glass (red dots) chamber.

### 4.3 Measurement with Positrons

Positron events were obtained with the secondary beam and the requirement of a signal in one of the two upstream Cerenkov counters. Data were collected at the 1, 2, 4, 8, and 16 GeV/c momentum sedttings. Figure 4 shows the distribution of the total number of hits in the calorimeter for the different energy settings and in comparison with the GEANT4 based simulation. Reasonable agreement between data and simulation is observed [13].

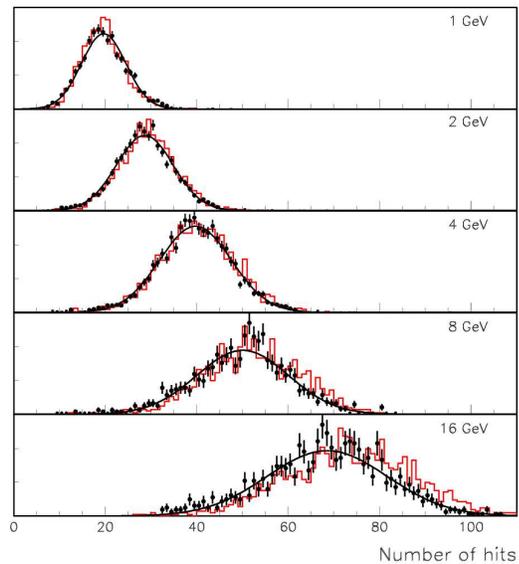

Figure 4: Distribution of the total number of hits in positron induced showers: data in black and simulation in red.

### 4.4 Measurements with Pions

Pion events were collected using the secondary beam and vetoing on signals from either of the upstream Cerenkov counters. Figure 5 shows the distribution of the number of hits for events selected to have initiated showering between the 1$^{st}$ and 2$^{nd}$ layer. The distributions were fit to two components: positron contamination and pion induced events as obtained from the GEANT4 based simulation. The parameters of the RPC simulation were obtained from the muon and positron shower data and no further tuning was applied to describe the pion data. Reasonable agreement between data and simulation is observed [15].

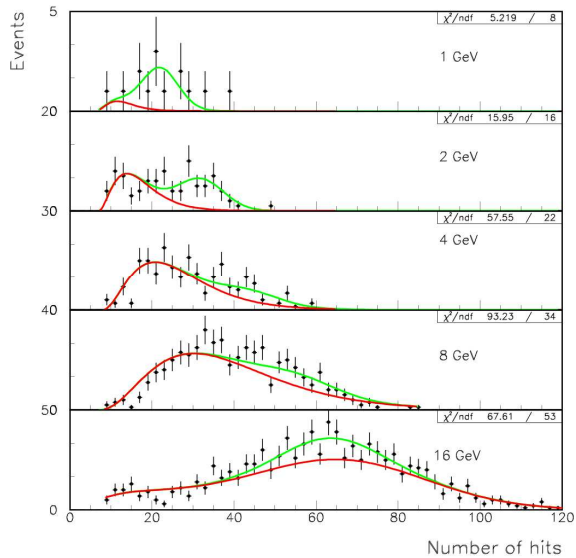

Figure 5: Distribution of the total number of hits in pion induced showers: data in black fit to the sum of two components (positron contamination in green and pion signal in red).

## 5  Simulation of the performance of a larger prototype calorimeter

Encouraged by the successful simulation of the pion shower data, the simulation tools were utilized to predict the performance of a larger prototype calorimeter with a front-face of 1×1 m$^2$ and 38 detector layers, corresponding to a depth of about four interaction lengths $\lambda_I$. The response was found to be approximately linear for pions with energies up to 25 GeV, Above this energy the response deviates from linearity due to overlapping effects in the 1×1 cm$^2$ pads. At 60 GeV the response is diminished by about 10 %.

Figure 6 shows the predicted resolution as function of pion energy. A fit of the points up to 28 GeV yields a stochastic term of the order of 58% with a vanishing constant term.

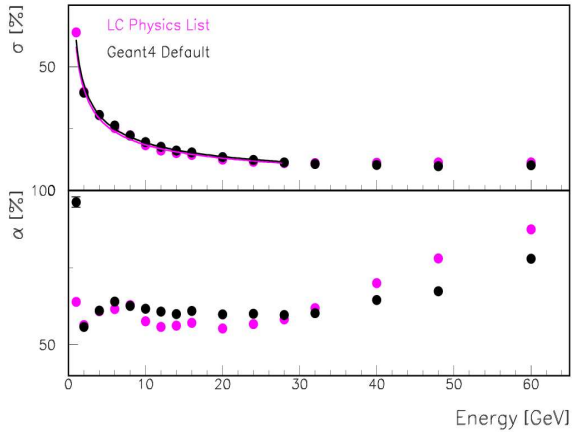

Figure 6: Top: Relative width obtained from Gaussian fits to the distribution of the number of hits versus pion energy. The lines are fits to the quadratic sum of a stochastic and a constant term. The simulations are based on the LC Physics list (magenta dots) and the Fast and Simple Physics list (black dots). Bottom: Stochastic terms (for energies up to 20 GeV) or Gaussian widths multiplied by $\sqrt{E}$ for energies above 20 GeV.

# 6  Acknowledgements

The author would like to thank the organizers for an impeccably organized workshop and for the opportunity to present the work of the DHCAL collaboration to the International Linear Collider community.